\documentclass[aps,pra,twocolumn]{revtex4-1}

\usepackage[T1]{fontenc}
\usepackage{graphicx,amsmath,amssymb,amsfonts,dsfont}
\usepackage{multirow,colortbl}
\usepackage[colorlinks=true,citecolor=blue,urlcolor=purple]{hyperref}
\usepackage{slashbox}

\def\ket#1{|#1\rangle}
\def\braket#1#2{\langle#1|#2\rangle}
\def\ketbra#1{|#1\rangle\langle#1|}
\def\bra#1{\langle#1|}
\def\mub#1#2{\ket{\varphi_{#2}^{#1}}}
\def\ub#1#2{\varphi_{#2}^{#1}\rangle}
\def\bmub#1#2{\bra{\varphi_{#2}^{#1}}}
\def\bmu#1#2{\langle\varphi_{#2}^{#1}}
\def\pmub#1#2{\ketbra{\varphi_{#2}^{#1}}}
\def\sectionaps#1{{\it#1}.~---}
\def\seb#1{{\color{black}#1}}
\newcommand{\calg}{\mathcal{G}}
\newcommand{\ii}{\mathrm{i}}
\newcommand{\calt}{\mathcal{T}}
\newcommand{\call}{\mathcal{L}}

\newcommand{\ff}{\mathbb{F}_d}
\newcommand{\gr}{\mathrm{GR}(4,r)}
\newcommand{\id}{\mathds{1}}

\newcommand{\vx}{{\vec{x}}}
\newcommand{\vl}{{\vec{l}}}
\newcommand\hb[2]{\genfrac{}{}{0pt}{}{#1}{#2}}
\newcommand{\vrj}{{\vec{j}}}
\newcommand{\rj}{j}
\newcommand{\cc}{\cellcolor[gray]{0.9}}
\renewcommand{\leq}{\leqslant}
\renewcommand{\geq}{\geqslant}

\newcommand\T{\rule{0pt}{3.5ex}}
\newcommand\B{\rule[-2ex]{0pt}{0pt}}
\DeclareMathOperator{\tr}{tr}
\DeclareMathOperator{\Tr}{Tr}

\makeatletter
\def\CT@@do@color{%
  \global\let\CT@do@color\relax
  \@tempdima\wd\z@
  \advance\@tempdima\@tempdimb
  \advance\@tempdima\@tempdimc
  \advance\@tempdimb0.9\tabcolsep
  \advance\@tempdimc\tabcolsep
  \advance\@tempdima2\tabcolsep
  \kern-\@tempdimb
  \leaders\vrule
  \hskip\@tempdima\@plus 1fill
  \kern-\@tempdimc
\hskip-\wd\z@ \@plus -1fill }
\makeatother

\begin{document}

\title{Quantifying measurement incompatibility of mutually unbiased bases}

\author{S\'ebastien Designolle,$^1$ Paul Skrzypczyk,$^2$ Florian Fr\"owis,$^1$ and Nicolas Brunner$^1$}
\affiliation{
  \makebox[0pt]{$^1$D\'epartement de Physique Appliqu\'ee, Universit\'e de Gen\`eve, 1211 Gen\`eve, Switzerland}\\
  \makebox[0pt]{$^2$H.H.~Wills Physics Laboratory, University of Bristol, Tyndall Avenue, Bristol, BS8 1TL, United Kingdom}
}

\date{\today}

\begin{abstract}
Quantum measurements based on mutually unbiased bases are commonly used in quantum information processing, as they are generally viewed as being maximally incompatible and complementary.
Here we quantify precisely the degree of incompatibility of mutually unbiased bases (MUB) using the notion of noise robustness.
Specifically, for sets of $k$ MUB in dimension $d$, we provide upper and lower bounds on this quantity.
Notably, we get a tight bound in several cases, in particular for complete sets of $k=d+1$ MUB (\seb{using the standard construction for $d$ being a prime power}).
On the way, we also derive a general upper bound on the noise robustness for an arbitrary set of quantum measurements.
Moreover, we prove the existence of sets of $k$ MUB that are operationally inequivalent, as they feature different noise robustness, and we provide a lower bound on the number of such inequivalent sets up to dimension 32.
Finally, we discuss applications of our results for Einstein-Podolsky-Rosen steering.
\end{abstract}

\maketitle

\sectionaps{Introduction}
Contrary to classical physics, different measurements in quantum mechanics can be incompatible, meaning that one cannot have access to their results simultaneously.
Incompatible measurements thus provide complementary information about a quantum system.
Motivated by the question of finding the measurements that are ``maximally incompatible'', Schwinger and others \cite{Sch60,Kra87,Iva81,Woo89} discussed the concept of mutually unbiased (bases) measurements.

Formally, in a complex Hilbert space of dimension $d$, two orthonormal bases $\{\ket{\varphi_a}\}_{a=1,\ldots, d}$ and $\{\ket{\psi_b}\}_{b=1,\ldots, d}$ are called \emph{mutually unbiased} if
\begin{equation}
  |\braket{\varphi_a}{\psi_{b}}|=\frac{1}{\sqrt{d}}
  \label{eqn:mub}
\end{equation}
for all $a$ and $b$.
That is, if a system is prepared in any eigenstate of one of the bases, then performing a measurement in the other basis gives a uniformly random outcome.
These bases can thus be considered ``maximally non-commutative'' and ``complementary'' \cite{Sch60}.

It is then natural to look for sets of $k$ measurements, such that all pairs are mutually unbiased \cite{Iva81}.
In the simplest case of qubits ($d=2$), there are three mutually unbiased bases (MUB) which are given by the eigenstates of the three Pauli observables.
In arbitrary dimension $d$, however, the construction of MUB is a difficult task.
It is proven that $k \leq d+1$ \cite{Ben06}, and an explicit construction of complete sets of $k=d+1$ MUB is only known when the dimension is a power of a prime, i.e., ${d=p^r}$ with $p$ prime and $r$ positive integer \cite{Woo89}.
A long-standing open problem is to determine the maximal number of MUB in the case $d=6$, which is conjectured to be $k=3$ \cite{Jam10,Bri08}.

More generally, MUB play a central role in quantum information processing \cite{Dur10}, and have been used in a wide range of applications such as quantum tomography \cite{Iva81,Woo89}, uncertainty relations \cite{Kra87,Maa88,Bal07}, quantum key distribution \cite{Bec03,Cer02}, quantum error correction \cite{Cal97}, as well as for witnessing entanglement \cite{Hua10,Spe12,Mac15,Reh13,Pau16,Erk17} and more general forms of quantum correlations \cite{Skr15,Sau17,Cos17}.
MUB also have strong links to other mathematical structures \cite{Pla06} such as finite projective planes \cite{San04,Hal08} or orthogonal Latin squares \cite{Pat09}.

Given the general significance of MUB, it is important to characterize their properties.
While MUB represent intuitively the most incompatible quantum measurements, the goal of the present work is to precisely quantify the degree of incompatibility of arbitrary sets of MUB.
As a measure of incompatibility we determine the noise robustness \cite{Hei15,Haa15,Cav16,Cav17}, namely the minimal amount of white noise required to make a given set of $k$ MUB in dimension $d$ jointly measurable \cite{Bus96,Bus10,Yu10,Pal11,Car12,Hei16,Uol16}, i.e., compatible.
We derive upper and lower bounds on this quantity for any $k$ and $d$.
Notably, we obtain tight bounds in many cases, in particular for $k=d$ and $k=d+1$ \seb{by using the standard construction of \cite{Woo89}} when $d$ is a prime power.
On the way, we also derive a general upper bound on the noise robustness for an arbitrary set of quantum measurements.

Moreover, these results highlight some interesting properties of MUB.
In particular, we find that there exist operationally inequivalent sets of MUB, in the sense that they feature a different noise robustness.
Lower bounds on the number of inequivalent sets are obtained for ${k \leq 8}$ and ${d \leq 32}$.
In fact, we observe that this phenomenon becomes generic in high dimensions.
Finally, our results also have direct implications for Einstein-Podolsky-Rosen steering \cite{Wis07}.
Exploiting the strong connection existing between joint measurability and steering \cite{Tul14,Uol14,Uol15}, we characterize the noise robustness of a broad class of entangled states in steering experiments.

\sectionaps{Preliminaries}
We consider sets of general quantum measurements, i.e., positive operator valued measures (POVMs).
A POVM is a collection of positive-semidefinite operators summing up to identity; given a state $\rho$ and a POVM $\{A_a\}_a$, the probability of getting outcome $a$ is then $p(a)=\tr A_a\rho$.
Our main focus will be to determine whether sets of POVMs (mostly noisy MUB) are compatible or not.
Note that the usual notion of commutativity, used for the case of projective measurements, is inadequate for general POVMs \cite{Kru87}.
Instead we use the notion of joint measurability \cite{Bus96,Hei16}.

Specifically a set of POVMs is jointly measurable if there exists a \emph{parent} POVM from which each POVM of the set can be recovered by taking the marginals.
This implies that, for any state $\rho$, the statistics of all POVMs in the original set can be recovered by first measuring the parent POVM, and then classically post-processing the result.
Formally, for a set of $k$ POVMs $\{\{A_{a|x}\}_a\}_{x=1,\ldots,k}$, joint measurability is ensured by the existence of a POVM $\{\calg_{\vrj[k]}\}_{\vrj[k]}$ such that
\begin{equation}
  \sum_{\hb{j_1,\ldots,j_{x-1}}{j_{x+1},\ldots,j_k}}\!\!\!\calg_{j_1,\ldots,j_{x-1},a,j_{x+1},\ldots,j_k}=\sum_{\vrj[k]}\delta_{\rj_x,a}\calg_{\vrj[k]}=A_{a|x}.
\end{equation}
Here and in the following, the notation $\vec{j}[k]$, often abbreviated $\vec{j}$ if $k$ is clear in the context, means $j_1,\ldots,j_k$.

Beyond this dichotomy of compatible vs incompatible, it is interesting to \emph{quantify} how incompatible a set of POVMs is.
A general way to do so consists in mixing each POVM of the set with white noise.
This defines a new set of noisy POVMs, where each POVM element is given by
\begin{equation}\label{noisyPOVM}
  A_{a|x}^\eta=\eta A_{a|x} + (1-\eta)\tr A_{a|x}\frac{\id}{d}.
\end{equation}
Physically, for rank-one projective measurements, this amounts to performing the measurement $A_{a|x}$ with probability $\eta$, and outputting at random with probability $1-\eta$.
By adding more and more noise to a set of incompatible POVMs, it is intuitive that it will eventually become jointly measurable.
Indeed, when $\eta=0$, only white noise remains so that joint measurability is ensured.
The critical parameter $\eta^*$ at which the transition occurs is the noise robustness, a meaningful incompatibility quantifier \cite{Hei15,Cav16,Cav17}.

\sectionaps{General upper bound}
First we consider a general set of $k$ POVMs $\{\{A_{a|x}\}_a\}_x$.
Its noise robustness $\eta^*$ can be expressed as the following semidefinite program (SDP) \cite{Hei15}; see also \cite{Wol09}.
\begin{align}\label{eqn:primal}
  \eta^*=\max_{\eta,\{\calg_\vrj\}_\vrj} &\quad \eta \nonumber \\
  \text{s.t.~} &\quad \sum_\vrj \delta_{\rj_x,a} \calg_\vrj = A_{a|x}^\eta \quad \forall a,x, \\
  &\quad \calg_\vrj \geq 0 \quad \forall \vrj, \quad \eta \leq 1. \nonumber
\end{align}
This formulation is well-known and has already been studied \emph{numerically}, even with MUB \cite{Bav17}.
Nonetheless, since we want \emph{analytical} results, we make use of a powerful tool used to study SDP, namely, duality theory.
Every SDP admits a dual program whose solution is greater than (weak duality) or equal to (strong duality) the primal one \cite{Boy04}.
In our case, the dual formulation of \eqref{eqn:primal} is
\begin{align}
  \eta^*=\min_{\{X_{a|x}\}_{a,x}}\!\!\! &\quad 1 +\tr\sum_{a,x}X_{a|x}A_{a|x} \label{eqn:dual}\\
  \text{s.t.~} &\quad  1 +\tr\sum_{a,x}X_{a|x}A_{a|x}\geq \frac{1}{d}\sum_{a,x}\tr A_{a|x}\tr X_{a|x},\nonumber \\
  &\quad \sum_{a,x}\delta_{\rj_x,a}X_{a|x} \geq 0 \quad \forall \vrj,\nonumber
\end{align}
where $X_{a|x}$ are new (dual) variables.
The equality with $\eta^*$ is ensured since strong duality holds in our case (see Appendix \ref{app:dual} for details).

Importantly, from Eq.~\eqref{eqn:dual} it is then clear that the value of ${1 +\tr\sum_{a,x}X_{a|x}A_{a|x}}$ corresponding to any $\{X_{a|x}\}_{a,x}$ that satisfies the constraints is an upper bound to $\eta^*$.
In Appendix \ref{app:dual}, we show that the following variables satisfy and saturate the constraints
\begin{equation}
  X_{a|x}=\frac{\frac{\lambda}{k}\id-A_{a|x}}{\sum\limits_{a',x'}\left[\tr A_{a'|x'}^2-\frac{1}{d}\left(\tr A_{a'|x'}\right)^2\right]}
  \label{eqn:fea}
\end{equation}
where
\begin{equation}
  \lambda=\max_{\vrj}\left\|S_{\vrj}\right\|_\infty\quad\mathrm{and}\quad S_\vrj=\sum_{x=1}^kA_{j_x|x}.
  \label{eqn:defx}
\end{equation}
This gives rise to the following bound on the noise robustness
\begin{equation}
  \eta^* \leq \frac{\lambda-\sum\limits_{a,x}\left(\frac{\tr A_{a|x}}{d}\right)^2}{\sum\limits_{a,x}\left[\frac{\tr A_{a|x}^2}{d}-\left(\frac{\tr A_{a|x}}{d}\right)^2\right]}=\eta_\mathrm{up}.
  \label{eqn:etauppovm}
\end{equation}

When $\{\{A_{a|x}\}_a\}_x$ are rank-one projective measurements, this further simplifies to
\begin{equation}
  \eta_\mathrm{up}=\frac{\lambda-\frac{k}{d}}{k-\frac{k}{d}}.
  \label{eqn:etaupmub}
\end{equation}

\sectionaps{Upper bound for MUB}
Notably, the bound \eqref{eqn:etaupmub} is also valid for projective measurements onto $k$ MUB.
Since there are $d^k$ (i.e., exponentially many) operators $S_\vrj$ to check in the definition \eqref{eqn:defx} of $\lambda$, this becomes computationally intractable very quickly.
A way to get a quick estimate of $\eta_\mathrm{up}$ is to use a bound on the norm of sums of projectors from Ref.~\cite{Skr15}.
In our case, thanks to the relation \eqref{eqn:mub}, we get ${\lambda\leq1+(k-1)/\sqrt{d}}$ which gives
\begin{equation}
  \label{eqn:upup}
  \eta_\mathrm{up}\leq\frac{\frac{\sqrt{d}}{k}+1}{\sqrt{d}+1}.
\end{equation}
This simple expression is however rarely tight.

Note that to derive the bound \eqref{eqn:upup}, the only assumption used is the unbiasedness \eqref{eqn:mub}.
Later, we also derive a lower bound based only on this hypothesis.
However, in general, this alone is not sufficient to fix the value of $\eta^*$.
Indeed, as will be shown below, inequivalent sets of MUB can have different $\eta^*$.
Thus to go further than only bounding $\eta^*$, we will use explicit sets of MUB.

\begin{table*}
  \caption{
  Noise robustness $\eta^*$ of sets of $k$ MUB in dimension $d\leq7$.
  For each case, we give the exact or approached values of $\eta^*$ and the upper bound $\eta_\mathrm{up}$.
  Instances for which the bound is tight, i.e., $\eta^*=\eta_\mathrm{up}$, are indicated by shaded cells, in particular, $k=2$, $k=d$ and $k=d+1$.
  Moreover, this shows the existence of operationally inequivalent sets of MUB, featuring different values of $\eta^*$.
  For instance, there are two inequivalent quadruplets for $d=7$, and $\eta^*=\eta_\mathrm{up}$ holds for one of them.
  For $d=6$ only three MUB could be constructed so far, but a bound could still be derived for $k=4$ (see Appendix \ref{app:up46}).
  }
  \begin{tabular}{|c|c|c|cc|cc|cc|cc|}
    \hline
    \backslashbox{$k$}{$d$} &                             2                             &                                        3                                        &        \multicolumn                                        {2}{c|}{4}                               &                 \multicolumn                                                     {2}{c|}{5}                                  &        \multicolumn                              {2}{c|}{6}                    &        \multicolumn                        {2}{c|}{7}                \\ \hline
                        &                 $\eta^*=\eta_\mathrm{up}$                 &                            $\eta^*=\eta_\mathrm{up}$                            &          $\eta^*$          &                           $\eta_\mathrm{up}$                           &      \hspace{0.6cm}             $\eta^*$     &                               $\eta_\mathrm{up}$                              &          $\eta^*$          &                 $\eta_\mathrm{up}$                &          $\eta^*$          &             $\eta_\mathrm{up}$            \\ \hline
             2          &           \cc $\frac{1}{\sqrt{2}}\approx0.7071$           &                     \cc $\frac{1+\sqrt{3}}{4}\approx0.6830$                     &        \multicolumn                            {2}{c|}{\cc $\frac23\approx0.6667$}                  &                 \multicolumn                                  {2}{c|}{\cc $\frac{3+\sqrt{5}}{8}\approx0.6545$}               &        \multicolumn          {2}{c|}{\cc $\frac{4+\sqrt{6}}{10}\approx0.6449$} &        \multicolumn          {2}{c|}{\cc $\frac{5+\sqrt{7}}{12}\approx0.6371$} \T\B\\ \hline
                        &                            \cc                            &                                       \cc                                       &                            &                                                                        &                 \multicolumn                                  {2}{c|}{\cc $\frac{1+\sqrt{5}}{6}\approx0.5393$}               &                            &                                                   &                            &                          \\
    \multirow{-2}{*}{3} & \cc \multirow{-2}{*}{~$\frac{1}{\sqrt{3}}\approx0.5774$~} &     \multirow{-2}{*}{\cc ~$\frac{\cos({\pi}/{18})}{\sqrt{3}}\approx0.5686$~}    & \multirow{-2}{*}{~0.5469~} &                       \multirow{-2}{*}{~0.5556~}                       &                 \multicolumn                   {2}{c|}{\cc ~~$\frac{13-\sqrt{5}+\sqrt{30 (5+\sqrt{5})}}{48}\approx0.5312$~~} & \multirow{-2}{*}{~0.5204~} &             \multirow{-2}{*}{~0.5254~}            & \multirow{-2}{*}{~0.5101~} &              \multirow{-2}{*}{~0.5154~}  \\ \hline
                        &                                                           &                                       \cc                                       &        \multicolumn                   {2}{c|}{                 \cc                        }         &                                              &                                                                               &                            &                                                   &        \multicolumn                   {2}{c|}{\cc 0.4516}         \\
    \multirow{-2}{*}{4} &                                                           & \multirow{-2}{*}{\cc $\frac{1+3 \sqrt{5}}{16\vphantom{\sqrt{5}}}\approx0.4818$} &        \multicolumn                      {2}{c|}{\multirow{-2}{*}{\cc $\frac12=0.5000$}}            & \hspace{0.5cm}      \multirow{-2}{*}{0.4615} &                            \multirow{-2}{*}{0.4616}                           &   \multirow{-2}{*}{?}      &       \multirow{-2}{*}{$\leq 0.4550$}               &          ~0.4436~          &              ~0.4488~                    \\ \hline
             5          &                                                           &                                                                                 &        \multicolumn          {2}{c|}{\cc $\frac{3+2 \sqrt{3}}{15\vphantom{\sqrt{3}}}\approx0.4309$} &                 \multicolumn                                                {2}{c|}{\cc 0.4179}                              &        \multicolumn                              {2}{c|}{?}                    &          ~0.4049~          &            ~0.4120~                    \T\B\\ \hline
             6          &                                                           &                                                                                 &                            &                                                                        &                 \multicolumn                                                {2}{c|}{\cc 0.3863}                              &        \multicolumn                              {2}{c|}{?}                    &          ~0.3754~          &            ~0.3867~                    \T\B\\ \hline
             7          &                                                           &                                                                                 &                            &                                                                        &                                              &                                                                               &        \multicolumn                              {2}{c|}{?}                    &        \multicolumn                       {2}{c|}{\cc 0.3685}     \T\B\\ \hline
             8          &                                                           &                                                                                 &                            &                                                                        &                                              &                                                                               &                            &                                                   &        \multicolumn                       {2}{c|}{\cc 0.3318}     \T\B\\ \hline
  \end{tabular}
  \label{tab:bigtab}
\end{table*}

\sectionaps{Tightness for specific MUB}
Here we exploit a specific implementation of MUB to analytically and numerically investigate the behavior of the noise robustness $\eta^*$ and the performance of the upper bound $\eta_\mathrm{up}$.
Since the construction of complete sets of MUB in prime power dimensions by Wootters and Fields \cite{Woo89} was reformulated in many equivalent ways \cite{Ban02,Kla04,God05,Dur10}, we choose different implementations depending on our needs.
We use the construction of Ref.~\cite{Dur10} for numerical purposes since it is easy to implement, and the one of Ref.~\cite{Kla04} when it comes to analytical results.

Table \ref{tab:bigtab} presents the solution $\eta^*$ of the SDP \eqref{eqn:dual} together with the upper bound $\eta_\mathrm{up}$ defined in Eq.~\eqref{eqn:etaupmub} for low dimensions $d\leq7$.
In some cases (e.g., triplets in dimension five and quadruplets in dimension seven), two solutions were obtained depending on the choice of the subset of MUB.
We discuss these inequivalent sets in more details below.

Notice that the equality $\eta^*=\eta_\mathrm{up}$ holds in a number of cases (shaded cells).
In particular, when $k=2$, $k=d$, and $k=d+1$, we prove this tightness analytically by providing an explicit parent POVM for $\{A_{a|1}^{\eta^*}\}_a,\ldots, \{A_{a|k}^{\eta^*}\}_a$.
It is given by the operators
\begin{equation}
  \calg_{\vrj} = \left\{\begin{array}{ll}
      \Pi_\vrj &\mathrm{if}\ \left\|S_{\vrj}\right\|_{\infty}\!\!=\lambda\\[0.5cm]
      0&\mathrm{otherwise,}
  \end{array}\right.
  \label{eqn:optmodgen}
\end{equation}
where $\Pi_\vrj$ is the projector onto the eigenspace of $S_\vrj$ associated with the maximum eigenvalue ($\lambda$ in that case).

For $k=2$, this was already known \cite{Car12,Uol16} and the above parent POVM indeed coincides with the one proposed in Sec.~IV of Ref.~\cite{Uol16}.

For $k=d$ and $k=d+1$, the proof of validity and optimality of this parent POVM \eqref{eqn:optmodgen} is more involved and consists of the following steps.
(i) From Appendix \ref{app:dual} we know that, as soon as $\calg_\vrj$ is a parent POVM for noisy MUB, our upper bound is tight.
(ii) We express $\calg_\vrj$ as $\lim\limits_{n\rightarrow\infty}\calg_\vrj^{(n)}$ where $\calg_\vrj^{(n)}=(S_\vrj/\lambda)^n$.
(iii) We prove the normalization of the $\calg_\vrj^{(n)}$, namely, $\sum_\vrj \calg_\vrj^{(n)}\propto\id$, from which the normalization of $\calg_\vrj$ is set.
(iv) We compute the marginals of $\calg_\vrj^{(n)}$.
This step is the only one in which the assumption $k=d$ or $k=d+1$ is used.
The complete proof can be found in Appendix \ref{app:wootters}.

We stress that, although the proof gives a fully analytical result --- in the sense that the noise robustness $\eta^*$ is formally an eigenvalue of a specific operator --- actually solving analytically this eigenvalue problem leads to the resolution of a polynomial equation whose explicit solution does not exist in general.
Apart from the case of two MUB in any dimension \cite{Uol16}, the cases in which we found such an explicit form are listed in Appendix \ref{app:ana}.
Additionally, there are special cases in which the upper bound is also reached.
This can be seen numerically either by comparing the result $\eta^*$ of the SDP \eqref{eqn:dual} with the value of $\eta_\mathrm{up}$ or by checking that the operators defined in Eq.\eqref{eqn:optmodgen} form a parent POVM for $\{A_{a|1}^{\eta^*}\}_a,\ldots, \{A_{a|k}^{\eta^*}\}_a$ (see Appendix \ref{app:ana} for details).

\sectionaps{Inequivalent sets of MUB}
When constructing sets of $k$ MUB in dimension $d$, there is some freedom.
In fact, it is known that (for certain $k$ and $d$) there exist sets of MUB that are inequivalent under unitaries, overall complex conjugation and other trivial operations like permutations or phase shifts \cite{Bri10}.
\seb{In the following we will simply refer to such sets as \emph{inequivalent}.}

Here, we go one step further, and show that there are sets of MUB that are \emph{operationally inequivalent}, in the sense that they feature different values of $\eta^*$.
For instance, this is the case for $k=3$ and $d=5$, where there are two inequivalent sets (see Table \ref{tab:bigtab}).
\seb{From the definition \eqref{eqn:dual}, it is clear that operationally inequivalent sets are necessarily inequivalent.
However, the converse does not hold as proven, e.g., by pairs of MUB in dimension four \cite{Bri10}.}
Note that in practice computing $\eta^*$ becomes quickly demanding.
\seb{Nevertheless we can obtain lower bounds on the number of sets featuring a different value of the upper bound $\eta_\mathrm{up}$.
In turn, this gives a lower bound on the number of inequivalent sets; indeed equivalent sets give the same  $\eta_\mathrm{up}$ (see Eq.~\eqref{eqn:etaupmub}).}
In Table \ref{tab:ineq} we give lower bounds on the number of inequivalent sets of MUB.
Interestingly, inequivalent sets seem to become \seb{more and more frequent} in high dimension (except when $d$ is a power of two).

\begin{table}[t!]
  \caption{
  Lower bound on the number of inequivalent sets of MUB.
  Shaded cells indicate operationally inequivalent sets (different values of $\eta^*$).
  When $k$ is greater than the number of MUB constructed in Ref.~\cite{Woo89}, the cell is left empty.
  Dimensions for which no inequivalent sets were found are not presented (e.g., dimensions 4, 6, 8, 32).
  }
  \begin{tabular}{|c|c|c|c|c|c|c|c|c|c|c|c|c|c|c|c|c|c|}
    \hline
    \backslashbox{$k$}{$d$} &   5   &   7   &   9   &   11  &   13  &   15  &   16  & 17 & 19 & 20 & 21 & 23 & 25 & 27 & 28 &  29 & 31 \\ \hline
           3       & \cc 2 & \cc 1 & \cc 2 & \cc 1 & \cc 2 & \cc 2 & \cc 1 &  2 &  1 &  2 &  2 &  1 &  2 &  1 &  1 &  2  &  1 \\ \hline
           4       & \cc 1 & \cc 2 & \cc 3 & \cc 2 & \cc 4 & \cc 1 &   1   &  4 &  4 &  1 &  1 &  4 &  3 &  2 &  2 &  6  &  6 \\ \hline
           5       & \cc 1 & \cc 1 & \cc 3 &   2   &   5   &       &   1   &  8 &  5 &  1 &    &  6 &  6 &  2 &  1 &  19 & 11 \\ \hline
           6       & \cc 1 & \cc 1 &   3   &   4   &   7   &       &   1   & 15 & 13 &    &    & 22 &  9 &  6 &    &  67 & 50 \\ \hline
           7       &       & \cc 1 &   2   &   2   &   10  &       &   1   & 20 & 18 &    &    & 32 & 38 &  9 &    & 145 & 92 \\ \hline
           8       &       & \cc 1 &   1   &   2   &   7   &       &   2   & 23 & 22 &    &    & 35 &  ? &  ? &    &  ?  &  ? \\ \hline
  \end{tabular}
  \label{tab:ineq}
\end{table}

\sectionaps{Lower bound for MUB}
Here we construct a very general parent POVM for noisy MUB using only the mutual unbiasedness~\eqref{eqn:mub} of the bases.
It is a generalization of the construction presented in Ref.~\cite{Uol16} for two MUB.

Let $\ket{\chi_\vrj^1}$ be defined iteratively by $\ket{\chi_{\rj_1}^1}=\mub{1}{\rj_1}$, the $j_1$-th vector of the first basis, and
\begin{equation}
  \ket{\chi_{\vrj[k]}^1}=\left(\id+\alpha_k\sqrt{d}A_{\rj_k|k}\right)\ket{\chi_{\vrj[k-1]}^1},
  \label{eqn:chilow}
\end{equation}
where $\alpha_i$ are positive coefficients introduced for later optimization.
Now let $\ket{\chi_\vrj^y}$ be defined similarly but with basis indices circularly shifted according to $y=1,\ldots, k$.
Specifically, $\ket{\chi_{\rj_1}^y}=\mub{y}{\rj_1}$, the $j_1$-th vector of the $y$-th basis, and, in the exponents of Eq.~\eqref{eqn:chilow}, 1 becomes $y$, 2 becomes $y+1$ (modulo $k$), etc.
Now a straightforward iterative proof shows that
\begin{equation}
  \calg_{\vrj[k]}=\sum_{y=1}^k\ketbra{\chi_{\vrj[k]}^y}
  \label{eqn:gjlow}
\end{equation}
is, up to normalization, a parent POVM for $\{A_{a|1}^{\eta_k}\}_a,\ldots, \{A_{a|k}^{\eta_k}\}_a$ where $\eta_k$ is defined recursively by $\eta_1=1$ and
\begin{equation}
  \eta_k=\frac{\left(2\alpha_k\sqrt{d}+d\right)(k-1)\eta_{k-1}+\left(2\alpha_k\sqrt{d}+\alpha_k^2d\right)}{k \left(2\alpha_k\sqrt{d}+\left(\alpha_k^2+1\right)d\right)}.
  \label{eqn:etalow}
\end{equation}
Then we can optimize over the coefficients $\alpha_2,\ldots,\alpha_k$ to get the highest possible noise parameter (see Appendix \ref{app:low} for details).
The best value achieved is denoted $\eta_\mathrm{low}$.
Since an explicit parent POVM of $k$ MUB with a noise parameter $\eta_\mathrm{low}$ was constructed, the noise robustness $\eta^*$ is indeed greater than $\eta_\mathrm{low}$.
While these bounds are only tight when $k=2$ or $d=2$, they are straightforward to compute and quite insightful.
For ${d\leq7}$, its approximated values are given in Table \ref{tab:low} in Appendix \ref{app:low}.

\sectionaps{Implications for EPR steering}
Our results also have implications for EPR steering, due to the intimate relation that exists with joint measurability \cite{Tul14,Uol14,Uol15}.
Specifically, our bounds on $\eta^*$ imply bounds on the noise robustness of certain entangled states for demonstrating steering.
Consider quantum states of the form
\begin{equation}
\rho^w_\psi = w\ket{\psi}\bra{\psi} + (1-w)\id/d \otimes \tr_A \ket{\psi}\bra{\psi},
\end{equation}
where $\ket{\psi}$ is an arbitrary pure entangled state of dimension $d \times d$.
It is interesting to determine the critical noise robustness $w^*$, i.e., the smallest value of $w$ such that $\rho^w$ demonstrates steering from the first party (Alice) to the second (Bob).
In general, $w^*$ depends on the set of measurements performed by Alice.
In the case she performs $k$ (noiseless) MUB measurements, we have that $w^* = \eta^*$, and hence all our results apply directly.
In the general case where Alice can perform all possible measurements, then we get the upper bound $w^* \leq \eta^*$.
We refer to Appendix \ref{app:epr} for details.

\sectionaps{Conclusion}
We discussed the problem of quantifying the measurement incompatibility of MUB.
We derived bounds on the noise robustness, which turn out to be tight in many cases, in particular for \seb{the standard construction of} complete sets of $k=d+1$ MUB \cite{Woo89}.
While our proof does not apply directly to other constructions of complete sets of MUB, we nevertheless believe that our bound is tight for any construction.
We also provided a general upper bound on the noise robustness for any set of POVMs.
It would be interesting to see how this bound performs for measurements that are not MUB, and whether one could find analytical solutions in other cases.

We showed the existence of operationally inequivalent sets of MUB, and provided lower bounds on their number.
We observed that inequivalent sets become more and more frequent in high dimensions.
Whether there exist operationally inequivalent sets of $k=d+1$ MUB remains a problem to be addressed.

Finally, our results have direct implications for EPR steering.
An interesting open question is whether complete sets of $d+1$ MUB are the most robust among all sets of $d+1$ measurements, as conjectured in Ref.~\cite{Bav17}.
In Appendix \ref{app:optmub2}, we give further support for this conjecture by proving it, in particular, for qubit projective measurements.
For general qubit measurements as well as for higher dimensions, this question is left open.

\sectionaps{Acknowledgments} Financial support by the Swiss National Science Foundation (Starting grant DIAQ, NCCR-QSIT), the Royal Society (URF UHQT), and European ERC-AG MEC is gratefully acknowledged.

\appendix

\section{Upper bound\texorpdfstring{\\}{} for arbitrary sets of POVMs}
\label{app:dual}

\subsection{Dual problem}

Here we derive the dual formulation of the SDP \eqref{eqn:primal} in the general case of arbitrary POVMs.
To that end, let us first write down the Lagrangian for this problem by introducing the Lagrange multipliers (dual variables) $\{Z_{\vrj}\}_{\vrj}$, $\{X_{a|x}\}_{a,x}$ and $\theta$,
\begin{equation}\label{eqn:lag}
\begin{aligned}
  \mathcal{L} =&\eta + \tr\sum_{\vrj}Z_{\vrj} \calg_\vrj + \theta(1-\eta)\\
  &-\tr\sum_{a,x} X_{a|x}\left(\sum_{\vrj} \delta_{\rj_x,a} \calg_\vrj - A_{a|x}^\eta\right).
\end{aligned}
\end{equation}
Notice that if we restrict our attention to dual variables that satisfy $Z_\vrj \geq 0$ and $\theta \geq 0$, then the last two terms of the first line of Eq.~\eqref{eqn:lag} are non-negative whenever the original (primal) variables satisfy the constraints of \eqref{eqn:primal}.
Similarly, the second line of Eq.~\eqref{eqn:lag} vanishes in this case, and we see that we have the inequality $\eta \leq \mathcal{L}$.

We can further restrict our attention to sets of dual variables which make the Lagrangian $\mathcal{L}$ independent of the primal variables $\{\mathcal{G}_\vrj\}_\vrj$ and $\eta$.
To see this, we first factorize the Lagrangian,
\begin{align}\label{eqn:lag2}
  \mathcal{L} =& \tr\sum_\vrj \calg_\vrj\left[Z_\vrj-\sum_{a,x}\delta_{\rj_x,a}X_{a|x}\right] \\
  &+ \eta\left[1 +\tr\sum_{a,x}X_{a|x}\left(A_{a|x} - \tr A_{a|x}\frac{\id}{d}\right) - \theta\right]\nonumber \\
  &+ \theta + \frac{1}{d}\sum_{a,x}\tr A_{a|x}\tr X_{a|x}.\nonumber
\end{align}
Thus, if we consider only dual variables that satisfy the additional constraints
\begin{align}
  \theta&=  1 +\tr\sum_{a,x}X_{a|x}\left(A_{a|x} - \tr A_{a|x}\frac{\id}{d}\right) \label{eqn:eqdual1}\\
  Z_\vrj&= \sum_{a,x}\delta_{\rj_x,a}X_{a|x} \quad \forall \vrj \label{eqn:eqdual2},
\end{align}
then the two square brackets in Eq.~\eqref{eqn:lag2} vanish, and we are left with
\begin{align}
  \mathcal{L} &= \theta + \frac{1}{d}\sum_{a,x}\tr A_{a|x}\tr X_{a|x} \nonumber \\
  &= 1 +\tr\sum_{a,x}X_{a|x}A_{a|x},
\end{align}
where the second line follows from the constraint \eqref{eqn:eqdual1}.
Thus, we arrive at an interesting situation, where for a particular set of dual variables, the Lagrangian $\call$, which by construction was bigger than $\eta$, is in fact independent of the primal variables.
We can therefore obtain the tightest bound on $\eta$ by minimizing the Lagrangian over this choice of dual variables, which is known as the dual problem, namely,
\begin{align}
  \min_{\hb{\{X_{a|x}\}_{a,x}}{\{Z_\vrj\}_\vrj,\theta}}\!\!\! &\quad1 +\tr\sum_{a,x}X_{a|x}A_{a|x} \\
  \text{s.t.~} &\quad\theta + \frac{1}{d}\sum_{a,x}\tr A_{a|x}\tr X_{a|x} = 1 +\tr\sum_{a,x}X_{a|x}A_{a|x},\nonumber  \\
  &\quad Z_\vrj= \sum_{a,x}\delta_{\rj_x,a}X_{a|x} \quad \forall \vrj, \nonumber \\
  &\quad Z_\vrj \geq 0 \quad \forall \vrj, \quad \theta \geq 0. \nonumber
\end{align}
Additionally, we can use the first and second constraints to solve for $Z_\vrj$ and $\theta$ (formally, they are referred to as ``slack variables''), which allows us to arrive at the final, simplified version of the dual problem which is given in Eq.~\eqref{eqn:dual} of the main text
\begin{align}
  \min_{\{X_{a|x}\}_{a,x}} &\quad 1 +\tr\sum_{a,x}X_{a|x}A_{a|x} \label{eqn:dualg} \\
  \text{s.t.~} &\quad  1 +\tr\sum_{a,x}X_{a|x}A_{a|x} \geq \frac{1}{d}\sum_{a,x}\tr A_{a|x}\tr X_{a|x},\nonumber \\
  &\quad \sum_{a,x}\delta_{\rj_x,a}X_{a|x} \geq 0 \quad \forall \vrj. \nonumber
\end{align}

Finally, there is a theorem, known as the \emph{strong duality theorem}, which is very powerful, and warrants the name ``dual problem'': it states that if one can find a solution to either the primal problem or the dual problem that is strictly feasible (i.e., one can find positive-definite operators, rather than just positive-semidefinite operators that satisfy all the constraints), then the value of the dual problem is \emph{equal} to the value of the primal problem.
In the present case, taking $X_{a|x} = \mu \id$, for $\mu > 0$, gives a strictly feasible solution to the dual, and hence strong duality holds.

\subsection{Ansatz solution}

One of the key uses of the dual is that any feasible solution to the dual provides an upper bound on the primal problem.
Let us make the following ansatz for the operators $X_{a|x}$, namely,
\begin{equation}
  X_{a|x} = \alpha \id - \beta A_{a|x}
\end{equation}
for some $\alpha$ and $\beta$ that need to be determined.
For this ansatz to satisfy the constraints of the dual \eqref{eqn:dualg}, we must have
\begin{align}
  1 &+\tr\sum_{a,x}X_{a|x}A_{a|x} - \frac{1}{d}\sum_{a,x}\tr A_{a|x}\tr X_{a|x} \nonumber \\
  &= 1-\beta\sum_{a,x}\left[\tr A_{a|x}^2-\frac{1}{d}\left(\tr A_{a|x}\right)^2\right] \geq 0,
  \label{eqn:firstcond}
\end{align}
and
\begin{align}
  \sum_{a,x}\delta_{\rj_x,a}X_{a|x} &= \sum_{a,x}\delta_{\rj_x,a} \left(\alpha \id - \beta A_{a|x}\right) \nonumber \\
  &=  k \alpha \id - \beta\sum_x A_{\rj_x|x} \geq 0.
  \label{eqn:secondcond}
\end{align}
The first condition \eqref{eqn:firstcond} is saturated if we pick
\begin{equation}
  \beta = \frac{1}{\sum\limits_{a,x}\left[\tr A_{a|x}^2-\frac{1}{d}\left(\tr A_{a|x}\right)^2\right]},
\end{equation}
while, if we define
\begin{equation}
  \lambda = \max_\vrj \left\| \sum_x A_{\rj_x|x}\right\|_\infty\!\! ,
\end{equation}
which is the largest spectral radius of all the operators $\sum_x A_{\rj_x|x}$, then
\begin{equation}
  \alpha = \frac{\beta \lambda}{k} = \frac{\frac{\lambda}{k}}{\sum\limits_{a,x}\left[\tr A_{a|x}^2-\frac{1}{d}\left(\tr A_{a|x}\right)^2\right]}
\end{equation}
satisfies the second condition \eqref{eqn:secondcond}, and cannot be improved.
Substituting these values for $\alpha$ and $\beta$ into the definition of $X_{a|x}$, we get that
\begin{equation}
  X_{a|x}=\frac{\frac{\lambda}{k}\id-A_{a|x}}{\sum\limits_{a',x'}\left[\tr A_{a'|x'}^2-\frac{1}{d}\left(\tr A_{a'|x'}\right)^2\right]}
  \label{eqn:feasible}
\end{equation}
is a feasible point for the problem \eqref{eqn:dualg}.
Thus we finally arrive at the bound on the robustness
\begin{equation}
  \eta \leq \frac{\lambda-\sum\limits_{a,x}\left(\frac{\tr A_{a|x}}{d}\right)^2}{\sum\limits_{a,x}\left[\frac{\tr A_{a|x}^2}{d}-\left(\frac{\tr A_{a|x}}{d}\right)^2\right]}.
  \label{eqn:etaupgen}
\end{equation}

\subsection{Educated guess for a parent POVM}
\label{subsec:opt}

Here we show that if a parent POVM for the noisy POVMs is given by the operators $\calg_\vrj$ defined by
\begin{equation}
  \calg_{\vrj} = \left\{\begin{array}{ll}
      \Pi_\vrj &\mathrm{if}\ \left\|\sum\limits_x A_{\rj_x|x}\right\|_{\infty}\!\!=\lambda\\[0.5cm]
      0&\mathrm{otherwise,}
  \end{array}\right.
  \label{eqn:optmodgen2}
\end{equation}
where $\Pi_\vrj$ is the projector onto the eigenspace of $\sum_x A_{\rj_x|x}$ associated with the eigenvalue $\lambda$, then the bound \eqref{eqn:etaupgen} is tight.
In that sense, these operators are quite natural to try when the bound \eqref{eqn:etaupgen} is reached.
This follows from Karush-Kuhn-Tucker conditions \cite{Boy04} but we derive it explicitly for pedagogical reasons.

Looking at Eq.~\eqref{eqn:lag}, it is clear that \emph{sufficient} conditions for the equality $\eta=\call$ to hold are
\begin{align}
                                   \theta &= 0\label{eqn:opt1}\\
                    \tr Z_\vrj\calg_\vrj &= 0\quad\forall\vrj\label{eqn:opt2}\\
  \sum_{\vrj} \delta_{\rj_x,a} \calg_\vrj &=  A_{a|x}^\eta\quad\forall a,x.\label{eqn:opt3}
\end{align}
It is easy to see that choosing the ansatz \eqref{eqn:feasible} together with the operators $\calg_\vrj$ defined in Eq.~\eqref{eqn:optmodgen2} fulfills \eqref{eqn:opt1} and \eqref{eqn:opt2}.
Importantly, the definition \eqref{eqn:optmodgen2} is designed to make \eqref{eqn:opt2} true.
Therefore, if the operators defined in Eq.~\eqref{eqn:optmodgen2} satisfy the constraint \eqref{eqn:opt3}, that is, form a parent POVM for $\{A_{a|1}^{\eta}\}_a,\ldots,\{A_{a|k}^{\eta}\}_a$, then all three condition are satisfied so that $\eta=\call$.
Since equality between primal and dual objective functions can only be true at their optimum, we know that the parent POVM $\calg_\vrj$ is, in that case, associated to the noise robustness $\eta^*$.

It is remarkable that this educated guess of $\calg_\vrj$ works for some sets of MUB, as stated in the main text and proven in Appendix \ref{app:wootters}.
We did not find any other set of measurement with this property.
Moreover, we found a set of MUB for which the bound \eqref{eqn:etaupgen} is tight without the operators \eqref{eqn:optmodgen2} being a parent POVM for the noisy MUB (see Appendix \ref{app:ana} for details).

\section{Quadruplets of MUB in dimension six}
\label{app:up46}

In dimension six, the number of MUB one can construct is unknown \cite{Jam10}, even though there are now a strong belief that no more than three can exist.
In this section, we nonetheless assume that four MUB do exist in dimension six.
We show that we can still derive a bound on $\lambda$ defined in Eq.~\eqref{eqn:etaupmub}.
The idea is to use the mutual unbiasedness \eqref{eqn:mub} to get some information on the characteristic polynomial of $S_{\vrj}$.

Given $k$ MUB $\{\mub{1}{j}\}_j,\ldots,\{\mub{k}{j}\}_j$ and the corresponding ${S_\vrj=\sum_x \pmub{x}{j_x}}$, it is straightforward to check the following equalities
\begin{align}
  \tr {S_{\vrj}}=&k\\
  \tr {S_{\vrj}^2}=&k \left(\frac{k-1}{d}+1\right)\\
  \tr {S_{\vrj}^3}=&k \left(3\frac{k-1}{d}+1\right)+\sigma_\vrj^{(3)}\\
  \tr {S_{\vrj}^4}=&k \left[6\frac{k-1}{d}+1+\left(\frac{k-1}{d}\right)^2\right]+4 \sigma_\vrj^{(3)}+\sigma_\vrj^{(4)},
\end{align}
where we have defined
\begin{equation}
  \sigma_\vrj^{(3)}=\sum_{\hb{a,b,c=1}{\neq}}^k\braket{\varphi_{\rj_a}^a}{\varphi_{\rj_b}^b}\braket{\varphi_{\rj_b}^b}{\varphi_{\rj_c}^c}\braket{\varphi_{\rj_c}^c}{\varphi_{\rj_a}^a}
\end{equation}
and
\begin{equation}
  \sigma_\vrj^{(4)}=\sum_{\hb{a,b,c,d=1}{\neq}}^k\braket{\varphi_{\rj_a}^a}{\varphi_{\rj_b}^b}\braket{\varphi_{\rj_b}^b}{\varphi_{\rj_c}^c}\braket{\varphi_{\rj_c}^c}{\varphi_{\rj_d}^d}\braket{\varphi_{\rj_d}^d}{\varphi_{\rj_a}^a}
\end{equation}
with the sums running over indices that take different values (pairwise).
Thanks to Newton's identities, we can express the characteristic polynomial of $S_{\vrj}$ using the traces of successive powers of $S_{\vrj}$.
When $\mathrm{rank}(S_{\vrj})=4$, it reads
\begin{align}
  X^{d-4}\Bigg\{&X^4-kX^3+\frac{k(k-1)}{2}\left(1-\frac1d\right)X^2\\
    &-\left[\frac{k(k-1)(k-2)}{6}\left(1-\frac3d\right)+\frac{\sigma_\vrj^{(3)}}{3}\right]X\nonumber\\
    &+\frac{k(k-1)(k-2)(k-3)}{24}\left(1-\frac6d\right)\nonumber\\
  &+\frac{k(k-1)^2(k-2)}{8d^2}+(k-3)\frac{\sigma_\vrj^{(3)}}{3}+\frac{\sigma_\vrj^{(4)}}{4}\Bigg\}.\nonumber
\end{align}

In the specific case of $k=4$ and $d=6$, this gives a simple expression for the characteristic polynomial of $S_{\vrj}$, namely,
\begin{equation}
  \frac{1}{4} \left[\left(2 X^2-4 X+1\right)^2-\sigma _4\right]-\frac{1}{3} \sigma _3 (X-1).
\end{equation}
Then it can be seen that its maximum root is an increasing function of both $\sigma_\vrj^{(3)}$ and $\sigma_\vrj^{(4)}$ so that we get an upper bound on $\lambda$ by taking their trivial upper bounds $|\sigma_\vrj^{(3)}|\leq24/d\sqrt{d}$ and $|\sigma_\vrj^{(4)}|\leq24/d^2$ achieved by the triangle inequality.
This bound is approximately $x\leq2.183$ so that we have ${\eta_\mathrm{up}\leq0.4550}$.

Then we can also compute the value of $\eta_\mathrm{low}$ since it does not require the explicit form of the MUB.
This leads to the following bounds for the value of $\eta^*$ in the case of hypothetical quadruplets of MUB in dimension six
\begin{equation}
  0.4175\leq\eta^*\leq0.4550.
\end{equation}

All this procedure also works for quadruplets of MUB in other dimensions in which only three MUB are known to exist.
For example, in dimension ten, it gives ${0.3864\leq\eta^*\leq0.4213}$.

\section{Analytical values}
\label{app:ana}

\begin{table}
  \caption{
  Analytical forms of the noise robustness $\eta^*$ of $k$ MUB.
  The cases $k=2$ or $d=2$ are not given since they were already known (see, e.g., \cite{Uol16}).
  Multiple values in one cell are due to the existence of inequivalent sets.
  In dimension nine, for $k=4$ and $k=6$, the value concerns only one of the three inequivalent sets (see Table \ref{tab:ineq}).
  Importantly, as explained in the main text, in some cases the equality $\eta^*=\eta_\mathrm{up}$ is only valid up to numerical precision.
  }
  \begin{tabular}{|c|c|c|c|c|c|}
    \hline
     \backslashbox{$k$}{$d$} &                         3                         &             4             &                                                  5                                                  &            8               &            9               \\ \hline
     \multirow{2}{*}{3}  & \multirow{2}{*}{$\frac{\cos (\pi/18)}{\sqrt{3}}$} &                           & $\frac{1+\sqrt{5}}{6}$ &                            &       $\frac{1}{2}$        \\
                         &                                                   &                           &       $\frac{13-\sqrt{5}+\sqrt{30 \left(5+\sqrt{5}\right)}}{48} $      &                            & $\frac{1+\cos(\pi/9)}{4}$  \\ \hline
              4          &             $\frac{1+3 \sqrt{5}}{16}$             &       $\frac{1}{2}$       &                                                                                                     &                            & $\frac{8+3\sqrt{3}}{32}$    \\ \hline
              5          &                                                   & $\frac{3+2 \sqrt{3}}{15}$ &                                                                                                     &                            &                            \\ \hline
              6          &                                                   &                           &                                                                                                     &                            & $\frac{3+\sqrt{7}}{16}$    \\ \hline
              9          &                                                   &                           &                                                                                                     & $\frac{3+2\sqrt{3}}{21}$   &                            \\ \hline
  \end{tabular}
  \label{tab:ana}
\end{table}

In Table \ref{tab:bigtab} of the main text, we give most of the analytical values we found for $\eta^*$.
Here we add further ones in Table \ref{tab:ana} together with explanations on their origins.

When the operators $\calg_\vrj$ defined in Eq.~\eqref{eqn:optmodgen} form a valid parent POVM for noisy versions of $\{A_{a|1}\}_a,\ldots,\{A_{a|k}\}_a$, we know from Appendix \ref{app:dual} that the bound $\eta_\mathrm{up}$ is tight.
This provides us with a simple sufficient criterion to get analytical values for $\eta^*$.
For $k=d$ and $k=d+1$, as claimed in the main text and proven in Appendix \ref{app:wootters}, the validity of this condition holds analytically.
Moreover, for one of the triplets in dimension five, namely, the one giving rise to ${\eta_\mathrm{up}=(1+\sqrt{5})/6}$, we could check it explicitly.
However, in the other cases, it could only be checked numerically thus the equality $\eta^*=\eta_\mathrm{up}$ is only valid up to numerical precision.

Similarly, we could also compare the numerical output $\eta^*$ of the SDP \eqref{eqn:dual} with the value of the upper bound $\eta_\mathrm{up}$.
Other instances of the equality $\eta^*=\eta_\mathrm{up}$ were found by this method, up to numerical precision.
For example, for triplets in dimension nine, the POVM \eqref{eqn:optmodgen} is not a parent POVM though the upper bound \eqref{eqn:etaupmub} is reached, up to numerical precision.

In Table \ref{tab:ana}, we do not give any value, e.g., for $k=5$ and $d=5$ though we claim in the main text that our method gives an analytical result in that case.
This is because in that case, $\eta^*$ is a root of a polynomial which we cannot solve explicitly.

Eventually, there are cases in which an analytic form is known but not reproduced here because it is too heavy.
For example, for $k=7$ and $d=7$, $\eta^*\approx0.3685$ is the largest root of ${56X^3-28X^2+1}$, which can be solved explicitly.

\section{Tightness of the upper bound\texorpdfstring{\\}{} for specific sets of MUB}
\label{app:wootters}

\subsection{A sequence converging to \texorpdfstring{$\calg_\vrj$}{Gj}}
\label{subsec:seq}

The expression of $\calg_\vrj$, though naturally emerging from the procedure of Appendix \ref{app:dual}, is very difficult to manipulate since the expression of the eigenvectors of $S_\vrj$ is unknown in general.
Here we express $\calg_\vrj$ as the limit of the sequence
\begin{equation}
  \calg_\vrj^{(n)}=\left(\frac{S_\vrj}{\lambda}\right)^n,
  \label{eqn:seq}
\end{equation}
where we recall from the definition \eqref{eqn:defx} that ${\lambda=\max_{\vrj}\|S_{\vrj}\|_\infty}$.
This can easily be seen by writing $S_\vrj/\lambda$ in a diagonal form.

The point of expressing $\calg_\vrj$ like this is that the elements $\calg_\vrj^{(n)}$ of the sequence are simpler to handle.
Thus we will prove our results on each $\calg_\vrj^{(n)}$ so that it will also hold for the limit by continuity.

Interestingly, for $n\leq4$, without any other assumption than the unbiasedness of the bases, $\calg_\vrj^{(n)}$ can be tediously proven to form, up to normalization, a parent POVM for $\{A_{a|1}^{\eta^{(n)}}\}_a,\ldots, \{A_{a|k}^{\eta^{(n)}}\}_a$ with ${\eta^{(1)}=1/k}$, ${\eta^{(2)}=\frac{d+2(k-1)}{k(d+k-1)}}$, ${\eta^{(3)}=\frac{d^2+5(k-1)d+3(k-1)(k-2)}{k[d^2+3(k-1)d+(k-1)(k-2)]}}$, and ${\eta^{(4)}=\frac{d^3+9(k-1)d^2+2(k-1)(7k-13)d+4(k-1)(k-2)(k-3)}{k[d^3+6(k-1)d^2+(k-1)(6k-11)d+(k-1)(k-2)(k-3)]}}$.

With some more effort, it is also possible to extend this to any $n$ using a specific form for the MUB and the assumption $k=d$ or $k=d+1$.
This is the goal of the next subsections.

\subsection{Odd prime power}
\label{subsec:odd}

In Ref.~\cite{Kla04} the following bases together with the computational one are proven to form a complete set of MUB in dimension $d=p^r$ where $p$ is an odd prime number
\begin{equation}
  \mub{x}{a}=\frac{1}{\sqrt{d}}\sum_{l\in\ff}\omega_p^{\Tr(xl^2+al)}\ket{l},
  \label{eqn:expodd}
\end{equation}
where the basis label $x$ and the vector label $a$ are in the Galois field $\ff$ with $d$ elements, $\{\ket{l}\}_{l\in\ff}$ is the computational basis, $\omega_p$ is a $p$-th root of the unity $\exp(2\mathrm{i}\pi/p)$, and $\Tr$ is the trace on $\ff$, defined by $\Tr a=a+a^p+\ldots+a^{p^{r-1}}$ so that it belongs to ${\mathds{F}_p=\mathds{Z}_p}$.

We recall some basic notions about Galois fields.
For further information, we refer the interested reader to Refs \cite{Dur10,Lid86}.
A Galois field is a finite set with two internal operations that have basically the same properties as the usual addition and multiplication: associativity, commutativity, existence of a unit element and of an inverse for all elements, and distributivity.
Having all these properties is very restrictive and only finite sets with a number of element $d=p^r$ with $p$ prime and $r$ positive integer are able to satisfy them.
This is why the construction only works in prime power dimensions.
The trace on Galois fields is simply a linear map from the abstract field to the set $0,\ldots, p-1$.

To keep notations simple, we will only consider in the following the case $k=d$ and all these MUB except the computational basis.
By treating the computational basis separately, the proof can be straightforwardly adapted to the other cases, namely, the other subsets of $d$ MUB and the complete set of $d+1$ MUB.

\subsubsection{Normalization}
\label{subsubsec:nor}

Here we prove that for all $n$, $\calg_\vrj^{(n)}$ is a POVM, up to normalization.
Since the positivity is immediate from the definition \eqref{eqn:seq}, we are left to show that these operators sum up to $\id$, up to normalization.
Interestingly the proof will also turn out to be valid for any $k$ and any subset of MUB.
By definition \eqref{eqn:defx} we have
\begin{align}
  \label{eqn:nor1}
  \sum_{\vrj[k]\in\ff}S_\vrj^n&=\sum_{\vrj}\left(\sum_{x\in\ff}\pmub{x}{\rj_x}\right)^n\\
  \label{eqn:nor2}
  &=\sum_{\vrj}\sum_{\vx[n]}\pmub{x_1}{\rj_{x_1}}\ldots\pmub{x_n}{\rj_{x_n}}
\end{align}
Then we choose a basis $\alpha\in\ff$ and we introduce a closure relation $\sum_{l\in\ff}\pmub{\alpha}{l}$ between all factors.
This trick is equivalent to projecting all vectors in the basis $\{\mub{\alpha}{l}\}_l$.
Each term of \eqref{eqn:nor2} then becomes
\begin{equation}
  \sum_{\vl[n+1]\in\ff}\!\!\!\mub{\alpha}{l_1}\bmu{\alpha}{l_1}\pmub{x_1}{\rj_{x_1}}\ldots\pmub{x_n}{\rj_{x_n}}\ub{\alpha}{l_{n+1}}\bmub{\alpha}{l_{n+1}},
\end{equation}
so that we can regroup the scalar products $\bmu{\alpha}{l_i}\pmub{x_i}{\rj_{x_i}}\ub{\alpha}{l_{i+1}}$ depending on the value of $x_i$ to perform the sum over $\vrj$, namely, the right hand side of \eqref{eqn:nor2} is
\begin{equation}
  \sum_{\vx,\vl}\left(\prod_{m\in\ff}\sum_{\rj_m}\prod_{\hb{i}{x_i=m}}\bmu{\alpha}{l_i}\pmub{x_i}{\rj_{x_i}}\ub{\alpha}{l_{i+1}}\right)\mub{\alpha}{l_1}\bmub{\alpha}{l_{n+1}}.
  \label{eqn:nor4}
\end{equation}
Using the definition \eqref{eqn:expodd} and quadratic Gauss sums on Galois fields \cite{Lid86}, when $x_i\neq\alpha$, we have
\begin{equation}
  \bmu{\alpha}{l_i}\pmub{x_i}{\rj_{x_i}}\ub{\alpha}{l_{i+1}}=\frac1d\omega_p^{\Tr(\mu[-(l_i^2-l_{i+1}^2)+2\rj_x(l_i-l_{i+1})])},
  \label{eqn:scalodd}
\end{equation}
where $\mu=2^{-2}(x_i-\alpha)^{-1}$.
Hence, thanks to the geometric sum $\sum_{x\in\ff}\omega_p^{\Tr(ax)}=d\delta_a^0$, this becomes
\begin{equation}
  \sum_{\rj_m}\prod_{\hb{i}{x_i=m}}\bmu{\alpha}{l_i}\pmub{x_i}{\rj_{x_i}}\ub{\alpha}{l_{i+1}}\propto\delta_{\!\!\!\sum\limits_{\hb{i}{x_i=m}}\!\!\!(l_i-l_{i+1})}^0
\end{equation}
where $\delta_x^y$ is the Kronecker delta in $\ff$ which is 1 if $x=y$ and 0 otherwise.
Combining all this, we notice that
\begin{equation}
  l_1-l_{n+1}=\sum_{m\in\ff}\sum_{\hb{i}{x_i=m}}(l_i-l_{i+1})=0
\end{equation}
so that $\sum_\vrj\calg_\vrj^{(n)}$ is diagonal in the basis $\{\mub{\alpha}{l}\}_l$.
Since this is true for all $\alpha$, it seems reasonable that $\sum_\vrj\calg_\vrj^{(n)}$ is a multiple of $\id$.
The following lemma formalizes this idea.

Let $M$ be an operator which is diagonal in two MUB denoted by $\{\mub{1}{i}\}_i$ and $\{\mub{2}{j}\}_j$, i.e., $M=\sum_im_i^{(1)}\pmub{1}{i}=\sum_jm_j^{(2)}\pmub{2}{j}$.
Then by decomposing $\mub{2}{j}$ in $\mub{1}{i}$ we get ${m_i^{(1)}=\sum_l\bmu{1}{i}\pmub{2}{l}\ub{1}{i}m_l^{(2)}}$ which is constant since the bases are mutually unbiased.

Hence the normalization of the previously introduced $\calg_\vrj^{(n)}$ is achievable.
Thus, by going to the limit, $\calg_\vrj$ can also be normalized.
Note that the proof provided here is valid for all $k$, which means that all the operators defined in \eqref{eqn:optmodgen} are always POVMs, up to normalization.

\subsubsection{Marginals}
\label{subsubsec:mar}

Here we compute the marginals of $\calg_\vrj^{(n)}$, i.e., its sum over $\vrj[k]$ with $\rj_\alpha$ fixed to $\gamma$, where we set $\alpha$ and $\gamma$ are in $\ff$.
The goal is to show that these marginals are of the form $\eta\pmub{\alpha}{\gamma}+(1-\eta)\id/d$, up to normalization.
The first steps are essentially the same as in the previous section: brute-force development of the power into many sums, injection of many closure relations to write the marginal in the basis $\{\mub{\alpha}{x}\}_x$, explicit evaluation of all the scalar products involved to get rid of the sum over $\vrj[k]$, and combination of the resulting Kronecker deltas to get the diagonality of the marginal.
Then, by relabeling the indices $\vx$ so that cumbersome coefficients vanish, this becomes
\begin{equation}
  \label{eqn:mar1}
  \sum_{\vrj[k]}\delta_{\rj_\alpha,\gamma}S_\vrj^n=\sum_{l_1\in\ff}\left(\sum_{\vx[n]\in\ff}\sum_{l_2\ldots l_n}c_{\vx,\vl}\right)\pmub{\alpha}{l_1}
\end{equation}
where the coefficients $c_{\vx,\vl}$ are defined to be
\begin{equation}
  \frac{d^{k-1}}{d^n}\prod_{m\in\ff}\delta_{\!\!\!\sum\limits_{\hb{i}{x_i=m}}\!\!\!(l_i-l_{i+1})}^0\prod_{\hb{i}{x_i=0}}d\delta_{l_i}^{\gamma}\delta_{l_{i+1}}^{\gamma}\prod_{i=1}^n\omega_p^{\Tr(x_i(l_i^2-l_{i+1}^2))}.
  \label{eqn:prod3}
\end{equation}
At this stage, we introduce, for a given $\vx$, the partition $X_1,\ldots,X_\tau$ of $[1,n]$ that naturally emerges when grouping the $x_i$ by their values.
With this, the sum over $\vx$ can be decomposed into a sum over the partitions of $[1,n]$ and the sums over the different values taken by the $x_i$ in the different sets of each partition.
Let $X_1,\ldots,X_\tau$ be a fixed partition of $[1,n]$.
We denote by $\alpha_\rho$ the common value of $x_i$ for $i\in X_\rho$.
Up to some manipulation, the computation of the coefficients of the diagonal expansion \eqref{eqn:mar1} leads to the evaluation of the following sum
\begin{equation}
  \sigma(\vec{\beta}[\epsilon])=\!\sum_{\alpha_1\in\ff^*}\omega_p^{\Tr(\alpha_1\beta_1)}\!\sum_{\hb{\alpha_2\in\ff^*}{\alpha_2\neq\alpha_1}}\!\omega_p^{\Tr(\alpha_2\beta_2)}\ldots\!\!\!\!\!\!\!\sum_{\hb{\alpha_\epsilon\in\ff^*}{\forall \rho<\epsilon,\alpha_\epsilon\neq\alpha_\rho}}\!\!\!\!\!\!\omega_p^{\Tr(\alpha_\epsilon\beta_\epsilon)}
\end{equation}
where $\beta_\rho=\sum_{i\in X_\rho}(l_i^2-l_{i+1}^2)$.
This is where the assumption on $k$ becomes a \emph{sufficient} condition to continue: if $k=d$ or $k=d+1$, the geometric sum $\sum_{\alpha_\epsilon\in\ff^*}\omega_p^{\Tr(\alpha_\epsilon\beta\epsilon)}=d\delta_{\beta_\epsilon}^0-1$ can be used to get a recursive expression of $\sigma(\vec{\beta}[\epsilon])$.
For example, when $k=d$, it reads
\begin{equation}
  \begin{aligned}
    \sigma(\vec{\beta}[\epsilon])=&\,(d\delta_{\beta_\epsilon}^0-1)\sigma(\vec{\beta}[\epsilon-1])\\&-\sum_{\rho=1}^{\epsilon-1}\sigma(\beta_1,\ldots,\beta_\rho+\beta_\epsilon,\ldots,\beta_{\epsilon-1}),
  \end{aligned}
\end{equation}
together with the initialization $\sigma(\beta_1)=d\delta_{\beta_1}^0-1$.
Using this, the coefficients of the diagonal expansion \eqref{eqn:mar1} become a complicated sum of cardinals of sets.
Fortunately, these cardinals are invariant when $l_1\neq \gamma$ varies so that it can be eventually seen that
\begin{equation}
  \begin{aligned}
    \sum_{\vrj[k]}\delta_{\rj_\alpha,\gamma}S_\vrj^n&=r\pmub{\alpha}{\gamma}+s\sum_{l_1\neq \gamma}\pmub{\alpha}{l_1}\\
    &=(r-s)\pmub{\alpha}{\gamma}+s\id,
  \end{aligned}
\end{equation}
with $r>s$.
The desired behavior, namely, that $\calg_\vrj$ forms a parent POVM for noisy versions of our MUB measurements, follows.

\subsection{Even prime power}
\label{subsec:even}

In Ref.~\cite{Kla04} the following bases together with the computational one are proven to form a complete set of MUB in dimension $d=2^r$
\begin{equation}
  \mub{x}{a}=\frac{1}{\sqrt{d}}\sum_{l\in\calt_r}\ii^{\Tr[(x+2a)l]}\ket{l},
  \label{eqn:expeven}
\end{equation}
where the basis label $x$ and the vector label $a$ are in the Teichm\"uller set $\calt_r$ of the Galois ring $\gr$, $\{\ket{l}\}_{l\in\calt_r}$ is the computational basis, and $\Tr$ is the trace on $\gr$.
We refer to Ref.~\cite{Wan12} for an introduction to Galois rings.

Then, by using Lemma 3 of Ref.~\cite{Car00} instead of quadratic Gauss sums, we can derive the equivalent of Eq.~\eqref{eqn:scalodd} when $x_i\neq\alpha$, namely,
\begin{equation}
  \bmu{\alpha}{l_i}\pmub{x_i}{\rj_{x_i}}\ub{\alpha}{l_{i+1}}=\frac1d\ii^{\Tr[t_i^{-2}(x_i-\alpha+2\rj_{x_i})(l_{i+1}-l_i)]},
  \label{eqn:scaleven}
\end{equation}
where $x_i-\alpha=t_i+2u_i$ with $t_i\in\calt_r^*$ and $u_i\in\calt_r$.
Then the proof is the same as in the odd case.
It is even simpler since there is no quadratic term in $l_i$ in the exponent of Eq.~\eqref{eqn:scaleven} so that the third product in Eq.~\eqref{eqn:prod3} disappears.

Thus the same result follows, namely, $\calg_\vrj$ is a parent POVM for noisy versions of our MUB measurements.

\section{Lower bound for MUB}
\label{app:low}

Here we give more details about the construction of the lower bound proposed in the main text, together with numerical values for ${d\leq7}$ in Table \ref{tab:low}.

We recall the procedure.
Using only the property of the bases to be mutually unbiased (see Eq.~\eqref{eqn:mub}), we can prove the operators $\calg_\vrj$ defined in Eq.~\eqref{eqn:gjlow} to form a parent POVM for noisy versions of our MUB measurements.
The corresponding noise parameter $\eta_k$ given in Eq.~\eqref{eqn:etalow} has some freedom intentionally introduced via the parameters $\alpha_i$, $2\leq i\leq k$.
As the construction provides, by Eq.~\eqref{eqn:primal}, a lower bound to $\eta^*$, we aim at maximizing the value of the function ${\eta_k(\alpha_2,\ldots,\alpha_k)}$.

The iterative definition \eqref{eqn:etalow} of $\eta_k(\eta_{k-1},\alpha_k)$ is increasing with $\eta_{k-1}$ since we choose positive $\alpha_k$.
Thus the optimization of the different parameters can clearly be made successively and independently.
We make it even more explicit in the following.
First we optimize ${\eta_2(\alpha_2)}$ over the single parameter $\alpha_2$.
By a simple derivative computation, we get the argument $\alpha_2^*=1$ for which the maximum $\eta_2^*$ of $\eta_2$ is reached.
Then we optimize ${\eta_3(\eta_2^*,\alpha_3)}$ over the single parameter $\alpha_3$.
This gives an argument
\begin{equation}
  \alpha_3^*=\frac{\sqrt{5 d+12 \sqrt{d}+8}-\sqrt{d}}{2 \left(\sqrt{d}+2\right)}
\end{equation}
giving rise to the maximum $\eta_3^*$ which we plug into the next step, and so on.

\begin{table}
  \caption{
  The lower bound $\eta_\mathrm{low}$ on the noise robustness $\eta^*$ we get by deriving an explicit parent POVM for $k$ MUB in dimension $d$ using only the mutual unbiasedness of the bases.
  }
  \begin{tabular}{|c|c|c|c|c|c|c|}
    \hline
    \backslashbox{$k$}{$d$} &    2   &    3   &    4   &    5   &    6   &   7     \\ \hline
           2       & 0.7071 & 0.6830 & 0.6667 & 0.6545 & 0.6449 & 0.6371  \\ \hline
           3       & 0.5774 & 0.5468 & 0.5263 & 0.5113 & 0.4996 & 0.4902  \\ \hline
           4       &        & 0.4672 & 0.4455 & 0.4297 & 0.4175 & 0.4076  \\ \hline
           5       &        &        & 0.3918 & 0.3758 & 0.3636 & 0.3537  \\ \hline
           6       &        &        &        & 0.3371 & 0.3250 & 0.3153  \\ \hline
           7       &        &        &        &        & 0.2958 & 0.2863  \\ \hline
           8       &        &        &        &        &        & 0.2634  \\ \hline
  \end{tabular}
  \label{tab:low}
\end{table}

\section{Implications for Einstein-Podolsky-Rosen steering}
\label{app:epr}

The results presented in the main text also have implications for EPR steering, due to the intimate relation that exists with joint measurability \cite{Uol14,Tul14,Uol15}.
In an EPR steering scenario, two parties called Alice and Bob share a quantum state, onto which Alice performs measurements, in order to ``steer'' the state of Bob's system.
In particular, if the shared state is $\rho^{AB}$, and Alice performs POVMs from the family $\{\{A_{a|x}\}_a\}_x$, then the sub-normalized states prepared for Bob are given by $\sigma_{a|x} = \tr_A[(A_{a|x}\otimes \id)\rho^{AB}]$, which occur with probability $p(a|x) = \tr \sigma_{a|x}$.
The collection of sub-normalized states $\{\{\sigma_{a|x}\}_a\}_x$ is known as an assemblage, and contains all of the observable data in an EPR steering scenario.
Steering is said to be demonstrated if this assemblage cannot be reproduced by a local-hidden-state (LHS) model, which takes the form of a collection of sub-normalized states $\{\sigma_\vrj\}_\vrj$ such that with probability $\tr \sigma_\vrj$, the state $\sigma_\vrj$ is sent to Bob and $\vrj$ is sent to Alice, who will announce $a = \rj_x$ as the outcome for the measurement labelled $x$.
The LHS model will reproduce the assemblage  $\{\{\sigma_{a|x}\}_a\}_x$ if $\sum_\vrj \delta_{\rj_x,a} \sigma_\vrj = \sigma_{a|x}$.

In Refs \cite{Tul14,Uol15} it was shown that joint measurability and EPR steering are equivalent in the following sense: any set of non-jointly-measurable measurements $\{\{A_{a|x}\}_a\}_x$, when used by Alice on any pure full-Schmidt-rank entangled state, $\ket{\psi} = \sum_i \sqrt{\lambda_i}\ket{i}\ket{i}$, where $\lambda_i > 0$ for all $i$, produces an assemblage $\{\{\sigma_{a|x}\}_a\}_x$ that demonstrates EPR steering.
That is, for pure full-Schmidt-rank states, it is necessary and sufficient to use non-jointly-measurable measurements to demonstrate EPR steering.

For any quantum state $\ket{\psi}$, a noisy version is given by ${\rho^w_\psi = w\ket{\psi}\bra{\psi} + (1-w)\id/d \otimes \tr_A \ket{\psi}\bra{\psi}}$.
For the maximally entangled state ${\ket{\Phi^+} = \sum_i \ket{i}\ket{i}}/\sqrt{d}$, then $\rho^w_{\Phi^+}$ is the isotropic state.
A basic question about EPR steering is to determine the critical noise $w^*$ below which a state $\rho^w$ becomes unsteerable (has an LHS model for all measurements) and above which it demonstrates steering.

The present analysis allows us to obtain the upper bound $w^* \leq \eta^*$.
In particular, for $\eta > \eta^*$ then projective MUB measurements are not jointly measurable, hence when Alice uses them on any full-Schmidt-rank state $\ket{\psi}$ the resulting assemblage demonstrates steering.
However, the identity
\begin{equation}
  \tr_A[(A_{a|x}^\eta\otimes \id)\ket{\psi}\bra{\psi}] = \tr_A[(A_{a|x}\otimes \id)\rho_\psi^\eta]
\end{equation}
shows that the exact same assemblage would arise from noiseless measurements on the noisy state $\rho_\psi^\eta$, which therefore also demonstrates steering.
Thus the values given in Table \ref{tab:bigtab} are bounds on the robustness of pure entangled states above which EPR steering can be demonstrated by using MUB measurements.

\section{Optimality of MUB'\texorpdfstring{\\}{} noise robustness in the qubit case}
\label{app:optmub2}

In this section we show that, among pairs (resp., triplets) of two-outcome unbiased qubit POVMs, the projective POVMs of two (resp., three) MUB reaches the minimal noise parameter $\eta^*=1/\sqrt{2}$ (resp., $\eta^*=1/\sqrt{3}$).
For pairs of POVM, this is actually already known \cite{Yu10}, but the proof is given to make the triplet-case clearer.
A two-outcome unbiased qubit POVM $A_i$ is given by
\begin{equation}
  A_{\mu_i|i}=\frac{\id+\mu_i\vec{a}_i\cdot\vec{\sigma}}{2},
\end{equation}
where $\mu_i$ can take the value $+1$ or $-1$.
It is said to be \emph{unbiased} because $\tr A_{+1|i}=\tr A_{-1|i}=1/2$.

For pairs of POVMs, we introduce the following operators
\begin{equation}
  G_{\mu_1\mu_2}=\frac{(1+\mu_1\mu_2Z)\id+\eta(\mu_1\vec{a}_1+\mu_2\vec{a}_2)\cdot\vec{\sigma}}{4},
\end{equation}
where $Z=1-\eta\|\vec{a}_1-\vec{a}_2\|$.
They obviously form a parent POVM of $A_1^\eta$ and $A_2^\eta$ as soon as they are all positive, which is the case whenever
\begin{equation}
  \eta\leq\eta_2^*=\frac{2}{\|\vec{a}_1+\vec{a}_2\|+\|\vec{a}_1-\vec{a}_2\|}.
\end{equation}
This is a well-known condition which is also sufficient \cite{Uol16}.
Through the Cauchy--Schwarz inequality, we get ${\eta_2^*\geq1/\sqrt{2}}$ with equality if and only if the two POVMs are projective MUB measurements, i.e., $\|\vec{a}_i\|=1$ and $\vec{a}_1\cdot \vec{a}_2=0$.

For triplets of POVMs, taking inspiration from Refs \cite{Yu13,Pal11}, we introduce
\begin{equation}
  \begin{aligned}
    G_{\mu_1\mu_2\mu_3}=\frac{1}{8}\bigg[&(1+\mu_2\mu_3Z_1+\mu_3\mu_1Z_2+\mu_1\mu_2Z_3)\id\\
    &+\eta(\mu_1\vec{a}_1+\mu_2\vec{a}_2+\mu_3\vec{a}_3)\cdot\vec{\sigma}\bigg],
  \end{aligned}
\end{equation}
where
\begin{align}
  Z_1=&1-\eta\frac{\|\vec{a}_2-\vec{a}_1-\vec{a}_3\|+\|\vec{a}_3-\vec{a}_1-\vec{a}_2\|}{2}\\
  Z_2=&1-\eta\frac{\|\vec{a}_1-\vec{a}_2-\vec{a}_3\|+\|\vec{a}_3-\vec{a}_1-\vec{a}_2\|}{2}\\
  Z_3=&1-\eta\frac{\|\vec{a}_1-\vec{a}_2-\vec{a}_3\|+\|\vec{a}_2-\vec{a}_1-\vec{a}_3\|}{2}.
\end{align}
They obviously form a parent POVM of $A_1^\eta$, $A_2^\eta$ and $A_3^\eta$ as soon as they are all positive, which is the case whenever
\begin{equation}
  \eta\leq\eta_3^*=\frac{4}{\Sigma},
\end{equation}
where
\begin{multline}
  \Sigma=\|\vec{a}_1+\vec{a}_2+\vec{a}_3\|+\|\vec{a}_1-\vec{a}_2-\vec{a}_3\|\\
  +\|\vec{a}_2-\vec{a}_1-\vec{a}_3\|+\|\vec{a}_3-\vec{a}_1-\vec{a}_2\|.
\end{multline}
This condition is only necessary \cite{Yu13} but it will be enough for our needs.
Indeed, through the Cauchy--Schwarz inequality, we get $\eta_3^*\geq1/\sqrt{3}$ with equality if and only if the two POVMs are projective MUB measurements, i.e., $\|\vec{a}_i\|=1$ and $\vec{a}_i\cdot \vec{a}_j=0$ for $i\neq j$.

In the case of pairs of POVMs, the result can even be extended to any pair of two-outcome qubit POVM.
In Ref.~\cite{Yu10}, the problem of joint measurement of any pair of two-outcome qubit POVM is indeed completely solved.
More specifically, an inequality involving the POVM parameters is given which is necessary and sufficient for their joint measurability.
For our needs, the sufficient inequality ${F_x^2+F_y^2 \geq 1}$ mentioned in Ref.~\cite{Yu10} is enough to ensure that the most robust to white noise pairs of two-outcome qubit POVM are projective measurements onto two MUB.

\end{document}